\documentclass[aps,pra,twocolumn,superscriptaddress,export]{revtex4-2} 
\usepackage[colorlinks=true,linkcolor=blue,citecolor=blue,urlcolor=blue]{hyperref}

\usepackage[utf8]{inputenc}
\usepackage[german, english]{babel}
\usepackage[dvipsnames]{xcolor}
\usepackage{bbm}
\usepackage{times}
\usepackage{tikz-cd}
\usepackage{xspace}
\hypersetup{colorlinks=true}
\usepackage{dcolumn}  
\usepackage{bm}        
\usepackage{bbold}
\usepackage{amsthm}
\usepackage{mathtools}
\usepackage{enumerate}
\usepackage[inline]{enumitem}
\usepackage{tikz}
\usetikzlibrary{shapes.geometric, arrows, shapes.multipart}
\usepackage{booktabs, tabularx}

\usepackage{blkarray}
\usepackage{xfrac}
\usepackage{stmaryrd}
\usepackage{qcircuit}
\usepackage{physics}

\usepackage{algorithm}
\usepackage{algpseudocode}
\algnewcommand\Input{\item[\textbf{Input:}]}%
\algnewcommand\Output{\item[\textbf{Output:}]}%

\usepackage[export]{adjustbox}
\usepackage{titlesec}
\usepackage{amsfonts}
\usepackage{amssymb}
\usepackage{braket}
\usepackage[normalem]{ulem}

\setenumerate{label={(\roman*)}}

\newcommand{\overbar}[1]{\mkern 1.5mu\overline{\mkern-1.5mu#1\mkern-1.5mu}\mkern 1.5mu}
\newcommand{\id}{I}

\newcommand{\CZ}{{\sf{CZ}}\xspace}

\newcommand{\MBQC}{{\text{MBQC}}\xspace}

\newcommand{\ins}{\mathcal{I}}
\newcommand{\outs}{\mathcal{O}}
\newcommand{\notouts}{\mathcal{M}}

\theoremstyle{remark}
\newtheorem{remark}{Remark}

\usepackage{amsthm}
\theoremstyle{plain}
\newtheorem*{theorem*}{Theorem}

\begin{document}

\title{Directly estimating the fidelity of measurement-based quantum computation}

\date{\today}

\author{David T. Stephen}
\affiliation{Quantinuum, 303 S Technology Ct, Broomfield, CO 80021, USA}
\author{Michael Foss-Feig}
\affiliation{Quantinuum, 303 S Technology Ct, Broomfield, CO 80021, USA}

\begin{abstract}

In measurement-based quantum computation (MBQC), quantum circuits are implemented using adaptive measurements on an entangled resource state. In practice, the resource state will always be prepared with some noise, and it is crucial to understand the effect of this noise on the operation of MBQC. Typically, one measures the fidelity of the noisy resource state with the assumption that a high fidelity state means a high fidelity computation. However, the precise relationship between these two fidelities is not known. Here, we derive an expression that equates the average fidelity of the MBQC output state to a certain correlation function evaluated on the noisy resource state. Using this expression, we show that state fidelity provides a tight lower bound on average MBQC fidelity. Conversely, we also find that state fidelity can greatly underestimate average MBQC fidelity,
implying that state fidelity is not a good indicator of MBQC performance in general. In response, we formulate an efficient method to directly estimate average MBQC fidelity by measuring the aforementioned correlation function. These results therefore improve our ability to characterize noisy resource states in quantum computers and benchmark MBQC performance.

\end{abstract}

\maketitle

\textit{Introduction.}---As quantum computers become increasingly sophisticated, it is essential to develop efficient tools to verify and benchmark their performance \cite{Gheorghiu2019,Eisert2020,kliesch2021theory}. Measurement-based quantum computation \cite{Raussendorf2001,raussendorf2003measurement}---a scheme of quantum computation that is driven by adaptive single-qubit measurements on an entangled multi-qubit \textit{resource state}---has unique advantages from the perspective of verification. In particular, the first step of MBQC, preparation of the resource state, can be efficiently verified by estimating the fidelity of the prepared state with respect to the ideal resource state \cite{Flammia2011,hayashi2015verifiable,Morimae2017hypergraph,Fujii2017verifiable,BermejoVega2018architectures}. The only difference between this verification and computation is the choice of single-qubit measurement bases. Therefore, assuming that measurement error is independent of the basis of measurement, one can benchmark the performance of MBQC by determining the fidelity of the prepared resource state. The possibility of efficiently verifying MBQC has lead to unique computational protocols, including verifiable quantum advantage  \cite{Gao2017,BermejoVega2018architectures,Haferkamp2020closing,Ringbauer2025,Fujii2016boundary,Fujii2017ising,Fujii2017verifiable,Miller2017supremacy,Miller2015,hayashi2019verifying} and verifiable blind quantum computation \cite{Barz2013,hayashi2015verifiable,Morimae2016,Fitzsimons2017unconditionally,Drmota2024verifiable,Li2023}. 

Despite the above aspirations, the precise relationship between resource state fidelity and MBQC fidelity remains poorly understood. It is standard to relate the two via the trace distance \cite{hayashi2015verifiable}: a large state fidelity implies a small trace distance between the noisy and ideal states, guaranteeing that any measurements performed on the two states, including those that drive MBQC, will return similar results. However, this is a very indirect way of assessing MBQC performance. Indeed, the trace distance is too stringent a metric since it bounds the total variation distance of all possible measurements, while in MBQC we are concerned only with a certain structured subset of all measurements. Furthermore, we only care about the quality of the final output state and not the distribution of measurement outcomes on non-output qubits. Accordingly, a state fidelity of $1-\epsilon$ only guarantees a trace distance $\sqrt{\epsilon}$ via the Fuchs-van de Graaf inequality \cite{nielsen2010quantum}, which is a relatively weak guarantee of MBQC performance. Therefore, it is desirable to find a more direct method of benchmarking MBQC.

In this letter, we establish a new framework for directly evaluating MBQC performance. We aim to benchmark a noisy resource state against an ideal stabilizer state that can be used to deterministically implement a certain parameterized set of circuits by adjusting the measurement bases in MBQC. In the case of universal resource states, this parameterized set maps efficiently to arbitrary quantum circuits. We then define the \textit{average MBQC fidelity} as the fidelity of the MBQC output when using the noisy resource state compared to the MBQC output when using the ideal state, averaged over all measurement bases. We show that this quantity has a remarkably simple expression as a certain correlation function evaluated on the noisy resource state. This expression allows us to prove tight bounds relating average MBQC fidelity and resource state fidelity. On one hand, we show that state fidelity is a direct lower bound of average MBQC fidelity, eliminating the quadratic reduction in precision when arguing via the trace distance. On the other hand, we find that state fidelity can greatly underestimate average MBQC fidelity, meaning that resource states with relatively low state fidelity can nevertheless lead to high-fidelity MBQC.

To address the loose relationship between state fidelity and MBQC fidelity, we show how to instead \textit{directly} estimate average MBQC fidelity using a sampling procedure. Importantly, this procedure is just as sample-efficient as estimating state fidelity. By directly estimating MBQC fidelity, we can gain more precise guarantees on the performance of MBQC with fewer samples. This result immediately suggests that many important applications of MBQC can be improved by supplanting the verification step based on state fidelity with the direct verification of MBQC fidelity. Overall, these results further bolster the superior verifiability of MBQC.


\vspace{3mm}
\textit{Measurement-based quantum computation.}---We analyze MBQC schemes that use stabilizer states as resources. We consider a certain class of stabilizer states that includes conventional resource states like cluster states \cite{raussendorf2003measurement} as well as more general resource states that have been the subject of recent investigation \cite{Herringer2023classificationof,herringer2024dualitystringcomputationalorder,Stephen2024universal}. A stabilizer state $\ket{\mathcal{S}}$ on $N$ qubits is defined by $N$ independent commuting stabilizers that generate a stabilizer group $\mathcal{G}$, such that $g\ket{\mathcal{S}} = \ket{\mathcal{S}}$ for all $g\in \mathcal{G}$. We let $\mathcal{V}$ denote the set of all qubits. Each stabilizer $g$ is a tensor product of Pauli operators $I$, $X$, $Y$, or $Z$. Throughout this letter, we refer to the stabilizer state $\ket{\mathcal{S}}$ as the \textit{ideal resource state}. It should be noted that most of the objects defined from here out will implicitly depend on $\ket{\mathcal{S}}$. We assume that no qubit in $\ket{\mathcal{S}}$ is disentangled from the rest, and we will soon impose more conditions on $\ket{\mathcal{S}}$ that ensure its ability to perform MBQC.

We wish to employ these stabilizer states, or noisy instances thereof, as resources for MBQC. 
To do this, we first need to define a set of \textit{output qubits} ${\mathcal{O}}\subset \mathcal{V}$ and the complementary set of \textit{measured qubits} $\notouts = {\mathcal{V}} \setminus {\mathcal{O}}$. 
MBQC proceeds by performing adaptive measurements on $\notouts$ such that the computational output is the state of the unmeasured output qubits $\mathcal{O}$, up to Pauli corrections that depend on the measurement outcomes. We assume without loss of generality that each qubit is measured in the $XY$-plane, meaning the basis $\ket{s^{\theta}} := U(\theta)\ket{s}$ for some angles $\theta$ where $s\in\{0,1\}$ with $\ket{0}$ and $\ket{1}$ the eigenstates of $Z$, and $U(\theta) = \exp (i\theta Z)H$ where $H=(X+Z)/\sqrt{2}$ \footnote{This assumption loses no generality since deterministic MBQC is only possible with measurements in the $XY$-, $XZ$, or $YZ$ plane, and we can always transform any $XZ$- or $YZ$-plane measurement into an $XY$-plane measurement with an appropriate local Clifford operator which can be absorbed into the resource state.}. The circuit executed by MBQC is then parameterized by the measurement angles $\bm{\theta} = \theta_1,\theta_2,\dots,\theta_M$ where $M=|\notouts|$ is the number of measured qubits. We also assume that the computational input is fixed in some state that is determined by the initial stabilizer group, but our results generalize to other input states.

We restrict our attention to stabilizer states that satisfy certain ``flow'' conditions which ensure the possibility of deterministic MBQC. While these conditions are usually stated for graph states \cite{hein2006entanglementgraphstatesapplications}  in terms of properties of the underlying graph \cite{Danos2006,browne2007generalized,Mhalla2008,Backens2021therebackagain}, they can be easily extended to arbitrary stabilizer states as follows.
We say a stabilizer state has flow if there exists a partial ordering $\prec$ of qubits such that, for every qubit $i\in \notouts$, there is an operator $R_i$ satisfying the following conditions:
\begin{enumerate*}
    \item $Z_iR_i$ is a stabilizer,
    \item $R_i$ acts as $X$ or $\id$ on all qubits in $\notouts$, and
    \item all qubits $j$ in the support of $R_i$ satisfy $i\prec j$
\end{enumerate*}
\footnote{More precisely, the notion of flow we use here has previously been called focused g-flow \cite{Mhalla2008,Backens2021therebackagain}}. 

To see why the flow conditions lead to deterministic MBQC, recall that each qubit $i\in\notouts$ is measured in the basis $\ket{s^{\theta_i}}$. The ordering $\prec$, which we refer to as the \textit{temporal order}, determines the order in which qubits are measured, where $i\prec j$ means $i$ must be measured before $j$. Because we have $\ket{1^\theta}  = Z\ket{0^\theta}$, obtaining one measurement outcome is equivalent to applying $Z$ and then obtaining the other. 
Using the fact that $Z_iR_i$ is a stabilizer, we can write $Z_i\ket{\mathcal{S}} = R_i\ket{\mathcal{S}}$ \footnote{This stabilizer exists even in the middle of the computation after some qubits have already been measured}. Since $R_i$ acts only on qubits $j$ such that $i\prec j$, we can undo the effect of inserting $Z_i$ by applying $R_i$ to the unmeasured qubits. For all measured qubits, $R_i$ acts as $I$ or $X$, and the latter can be absorbed into the measurement by flipping the measurement angle $\theta\rightarrow-\theta$. The Pauli action of $R_i$ on the output qubits needs to be actively corrected or taken care of with post-processing of the final measurement outcomes. Performing these adaptive measurements and Pauli corrections results in a final output state that is independent of the measurement outcomes.


To aid the upcoming analysis, we show how to reformulate MBQC in an equivalent way that does not require adaptive measurements, or even measurements at all, leading to the oxymoronic ``measurement-free measurement-based quantum computation'' (akin to measurement-free quantum error correction \cite{heussen2024measurement}). This reformulation serves as a technical tool that allows us to easily compute the MBQC output state for any resource state.

\begin{figure}
\centering
\mbox{
\Qcircuit @C=0.5em @R=0.5em 
    {
    \lstick{} & \gate{U(\theta_1)} & \ctrl{1}  & \qw & \qw & \qw & \qw & \qw & \qw  \\ 
     \lstick{} & \qw & \multigate{5}{R_1} &  \gate{U(\theta_2)} & \ctrl{1} & \qw & \qw & \qw & \qw  \\ 
      \lstick{} & \cds{2}{\vdots} & \ghost{R_1} & \cds{2}{\vdots} & \multigate{4}{R_2}  & \cds{2}{\ddots}  & \qw & \qw & \qw \\ 
     \lstick{} & & & & & & & & & \\
     \lstick{} & \qw  & \ghost{R_1} & \qw & \ghost{R_2}  & \qw  & \gate{U(\theta_M)} & \ctrl{1} & \qw  \inputgroupv{1}{5}{1em}{3em}{\notouts} \\ 
     \lstick{} & \qw  & \ghost{R_1} & \qw & \ghost{R_2} & \qw  & \qw & \multigate{1}{R_M} & \qw \\
     \lstick{} & \qw  & \ghost{R_1} & \qw & \ghost{R_2} & \qw  & \qw & \ghost{R_M} & \qw \inputgroupv{6}{7}{1em}{1em}{\outs}
    }
    }
    \caption{Illustration of the measurement-free MBQC operator $\Gamma(\bm{\theta})$ for an arbitrary stabilizer state (here with $|\outs| = 2$). The qubits are arranged from top to bottom according to the ordering defined by $\prec$.}
    \label{fig:mfmbqc}
\end{figure}
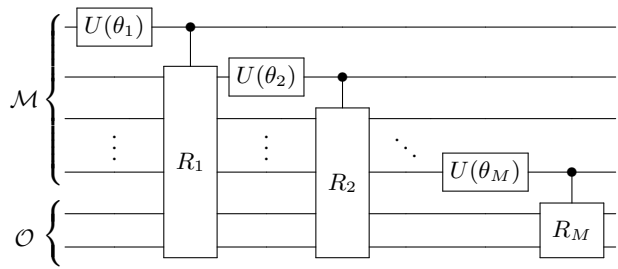

We begin with the observation that measuring a qubit in the $Z$-basis and applying a correction operator based on the measurement outcome is equivalent to first applying an appropriate controlled unitary from the to-be-measured qubit to the corrected qubits and then measuring. In the adaptive measurement scheme described above, we should apply $\mathsf{CR}_i$ before measuring qubit $i$ in the $Z$-basis (and after applying $U(\theta)$ to rotate the measurement into the $XY$-plane) where 
$
    \mathsf{CR}_i = \ketbra{0}{0}_i\otimes \id + \ketbra{1}{1}_i\otimes R_i
$
is the controlled-$R_i$ operator.
In this way, performing adaptive measurements on the resource state is equivalent to applying the unitary $\Gamma(\bm{\theta})$ and then measuring in the $Z$-basis without adaptivity, where,
\begin{equation} \label{eq:gamma}
    \Gamma(\bm{\theta}) = \prod_{i\in\notouts}  \mathsf{CR}_i U_i(\theta_i). 
\end{equation}
Therein, the product is taken according to the temporal order defined by $\prec$. This circuit is shown in  Fig.~\ref{fig:mfmbqc}.
For the ideal resource state, each measurement outcome of the qubits $i\in\notouts$ occurs with equal probability \footnote{This can be seen to result from the existence of the stabilizer $Z_iR_i$.}, and determinism implies that we get the same output for all measurement outcomes. This implies,
\begin{equation} \label{eq:psi_from_gamma}
    \ketbra{\bm{s}} \Gamma(\bm{\theta})\ket{\mathcal{S}} = \frac{1}{\sqrt{2^M}}\ket{\bm{s}}\otimes \ket{\psi(\bm{\theta})},
\end{equation}
for all measurement outcomes $\bm{s} = s_1,\dots,s_M$ where $\ket{\psi(\bm{\theta})}$ is called the \textit{ideal output state} corresponding to the measurement pattern $\bm{\theta}$. Therefore, applying the unitary operator $\Gamma(\bm{\theta})$ to $\ket{\mathcal{S}}$ already implements the MBQC without requiring measurements, in that the measured qubits are disentangled into the $\ket{+}$ state and the output qubits are mapped to $\ket{\psi(\bm{\theta})}$, hence the name measurement-free MBQC. We emphasize that measurement-free MBQC is only a technical tool, as the additional gates making up $\Gamma(\bm{\theta})$ would contribute unnecessary error compared to the adaptive measurement procedure.

\vspace{3mm}
\textit{Characterizing noisy resource states.}---Now we consider some other resource state $\rho$, which will in practice be a noisy version of the ideal resource state, but in principle can be any state defined on the same Hilbert space. We wish to directly characterize the performance of $\rho$ as a resource for MBQC. 
Suppose we perform the exact same pattern of adaptive measurements and corrections on $\rho$ as if it was the ideal resource state. Then, the output qubits will end up in some \textit{noisy output state} $\sigma_{\bm{s}}(\bm{\theta})$ which now generally depends on the measurement outcomes $\bm{s} = s_1,\dots,s_{M}$ with $s_i=0,1$. We define the MBQC fidelity as the fidelity of the noisy output state with respect to the ideal output state, averaged over measurement outcomes,
\begin{equation} \label{eq:mbqc_fidelity}
    F_\MBQC(\rho|\bm{\theta}) = \sum_{\bm{s}} p_{\bm{s}} \bra{\psi(\bm{\theta})}\sigma_{\bm{s}}(\bm{\theta})\ket{\psi(\bm{\theta})},
\end{equation}
where $p_{\bm{s}}$ is the probability of obtaining $\bm{s}$. We can further average over the measurement angles $\bm{\theta}$ to obtain the average MBQC fidelity,
\begin{equation} \label{eq:av_mbqc_fidelity}
    \overbar{F}_\MBQC(\rho) = \frac{1}{(2\pi)^{M}}\int F_\MBQC(\rho|\bm{\theta}) \ d\bm{\theta}
\end{equation}
where the integral is over all values of $\theta_i$ between $0$ and $2\pi$. The quantity $\overbar{F}_\MBQC(\rho)$ represents the average fidelity of all MBQCs implementable by performing $XY$-plane measurements on the resource state $\ket{\mathcal{S}}$.

It turns out that $\overbar{F}_\MBQC(\rho)$ has a remarkably simple expression for any resource state.
In order to succinctly describe the noisy output states, we use the measurement-free MBQC operator $\Gamma(\bm{\theta})$, which gives, 
\begin{equation}
    \sigma_{\bm{s}}(\bm{\theta}) \equiv \frac{1}{p_{\bm{s}}}\bra{\bm{s}} \Gamma(\bm{\theta}) \rho \Gamma(\bm{\theta})^\dagger \ket{\bm{s}} 
\end{equation}
Now we can insert this into Eq.~\eqref{eq:mbqc_fidelity} to obtain,
\begin{align}
    &F_\MBQC(\rho|\bm{\theta}) = \sum_{\bm{s}} \bra{\psi(\bm{\theta})}\bra{\bm{s}} \Gamma(\bm{\theta}) \rho \Gamma(\bm{\theta})^\dagger \ket{\bm{s}}\ket{\psi(\bm{\theta})} \nonumber \\
    &= {2^M}\sum_{\bm{s}} \bra{\mathcal{S}}\Gamma(\bm{\theta})^\dagger \ketbra{\bm{s}} \Gamma(\bm{\theta}) \rho \Gamma(\bm{\theta})^\dagger \ketbra{\bm{s}} \Gamma(\bm{\theta})\ket{\mathcal{S}} \nonumber \\
    &= \tr \rho \Omega(\bm{\theta})
\end{align}
where we used Eq.~\eqref{eq:psi_from_gamma} in the second line, and in the third line we used the cyclic property of the trace and defined,
\begin{equation} \label{eq:omega_theta}
    \Omega(\bm{\theta}) = {2^M}\sum_{\bm{s}} \Gamma(\bm{\theta})^\dagger \ketbra{\bm{s}} \Gamma(\bm{\theta})\ketbra{\mathcal{S}}\Gamma(\bm{\theta})^\dagger \ketbra{\bm{s}} \Gamma(\bm{\theta}). 
\end{equation}
Thus, the fidelity of the MBQC parameterized by the measurement angles $\bm{\theta}$ can be expressed in terms of the expectation value of the operator $\Omega(\bm{\theta})$ evaluated on the resource state. This generalizes the result of Ref.~\cite{chung2009characterizing} to arbitrary resource states and arbitrary measurement bases. 

In Appendix \ref{app:omega}, we show that integrating $\Omega(\bm{\theta})$ over all measurement angles $\bm{\theta}$ gives,
\begin{equation} \label{eq:mbqc_corr}
    \overbar{F}_\MBQC(\rho) = \tr \rho\Omega
\end{equation}
where,
\begin{equation} \label{eq:omega}
    \Omega = \frac{1}{2^{|\mathcal{O}|}}\sum_{g\in\mathcal{G}^{XY}} \frac{1}{2^{w_{\notouts}(g)}} g
\end{equation}
where $\mathcal{G}^{XY}\subset \mathcal{G}$ is the subset of all stabilizers that act as only $I$, $X$, or $Y$ on all measured qubits and $w_{\notouts}(g)$ is the weight of $g\in \mathcal{G}^{XY}$ [the number of qubits on which it acts non-trivially] restricted to $\notouts$. Eqs.~\eqref{eq:mbqc_corr} and \eqref{eq:omega} constitute the main result of our work.

As a simple example of Eq.~\eqref{eq:omega}, we consider the 1D cluster state on $N$ qubits whose stabilizer group is generated by $X_1Z_2$, $Z_{i-1}X_{i}Z_{i+1}$ for $1<i<N$, and $Z_{N-1}X_N$. 
Taking $\outs = \{N\}$, the correlation functions $\Omega$ for $N=2,3$ read,
\begin{equation} \label{eq:omega23}
\begin{aligned}
    \Omega_{N=2} &= \frac{1}{2}\id + \frac{1}{4}XZ + \frac{1}{4}YY \\
    \Omega_{N=3} &= \frac{1}{2}\id + \frac{1}{4}XIX + \frac{1}{8}(-YXY) + \frac{1}{8}YYZ.
\end{aligned}
\end{equation}
We will return to this example in the next section.

We make a few preliminary remarks about this result. First, as shown in Appendix \ref{app:spectrum}, the coefficients of each term in the sum add up to 1 and the stabilizers $g\in\mathcal{G}^{XY}$ generate all of $\mathcal{G}$. Therefore, $\overbar{F}_\MBQC(\rho) = 1$ if and only if $\rho = \ketbra{\mathcal{S}}$, meaning that perfect MBQC is possible only with the ideal resource state. We should compare this result with other seemingly contradictory results in the literature which show that MBQC performance is robust to perturbations respecting certain symmetries \cite{doherty2009identifying,Bartlett2010renormalization,Miyake2010,Miller2015,Else2012,Stephen2017,Stephen2019,Raussendorf2019,Daniel2019,Devakul2018mbqc,Raussendorf2023measurementbased}. These works each modify the MBQC protocol, by, for example, using measurements that drive the perturbed state back to the ideal state \cite{Bartlett2010renormalization,Miller2015}, changing the measurement bases depending on the perturbation \cite{Stephen2017,Raussendorf2023measurementbased}, or only considering certain measurement patterns \cite{doherty2009identifying,chung2009characterizing,Else2012}. Here, we assume that the perturbed state is measured exactly as if it was an ideal state, and we average over all measurement patterns, which rules out the above protocols.

Second, it is easy to modify $\Omega$ to capture the scenario where some qubits are always measured in the $X$ basis or always measured in the $Y$ basis. This is relevant if one wants to determine the average fidelity of MBQC circuits with parameterized single-qubit rotations mixed in with fixed patterns of Clifford gates. To achieve this, we simply restrict the sum in Eq.~\eqref{eq:omega} to include only those stabilizers that are compatible with the fixed measurements and then re-weight each non-zero stabilizer appropriately, see Appendix \ref{app:omega} for details. For example, suppose we take a 1D cluster state of length $N=9$ (the extension to larger $N$ is straightforward) and we measure all qubits $i\in\notouts$ in the $X$ basis, except for the middle qubit which we measure in the $XY$-plane. After averaging over $XY$-plane measurement basis, $\overbar{F}_\MBQC$ is equal to the expectation value of,
\begin{equation}
    \Omega = \frac{1}{2}\id + \frac{1}{4}XIXIXIXIX - \frac{1}{4} XIXI\,YXXXY.
\end{equation}
The product of the second and third terms is equal to $IIIIZXIXZ$ which is an example of a string order parameter for the 1D cluster state \cite{Pollmann2012a,Raussendorf2023measurementbased}. Therefore, our results are consistent with previous results that have identified the importance of string order in the performance of MBQC \cite{doherty2009identifying,chung2009characterizing,Raussendorf2023measurementbased, Masui2025}    . In this case, there \textit{are} perturbations to the ideal resource state that will preserve the fidelity of this class of MBQC circuits, namely those that commute with $\Omega$.

Finally, if we sum each $\theta_i$ over all angles that are a multiple of $\pi/4$ in Eq.~\eqref{eq:av_mbqc_fidelity}, rather than integrating over all angles, we obtain the same result as in Eq.~\eqref{eq:omega} (see Appendix \ref{app:omega}). This corresponds to measuring each qubit in the $\pm X$ or $\pm Y$ basis, which implements a Clifford circuit in MBQC. Therefore, we obtain the result that {the average fidelity over all circuits implementable by a given resource state is the same as the average fidelity over all \textit{Clifford} circuits implementable by that resource state.} This can also be seen directly from the fact that Eq.~\eqref{eq:omega_theta} is quadratic in matrix elements of $U(\theta)$ and $U(\theta)^{\dagger}$ and that the gate set $e^{i n\frac{\pi}{4} Z}H$ for $n=0,1,2,3$ forms a unitary 2-design for the gate set $U(\theta)$, akin to previous  relations between Haar-random circuits and random Clifford circuits \cite{Fisher2023random}.



\vspace{3mm}
\textit{Relationship between state and MBQC fidelities.}---Unsurprisingly, the average MBQC fidelity turns out to be closely related to state fidelity of $\rho$ with the ideal resource state,
\begin{equation} \label{eq:state_fidelity}
    F_S(\rho) = \bra{\mathcal{S}}\rho\ket{\mathcal{S}} = \frac{1}{2^{|\mathcal{V}|}}\sum_{g\in \mathcal{G}} \tr\rho g,
\end{equation}
where in the second equation we wrote the projector $\ketbra{\mathcal{S}} = \frac{1}{2^{|\mathcal{V}|}}\sum_{g\in \mathcal{G}} g$. 
Observe that both state fidelity \eqref{eq:state_fidelity} and average MBQC fidelity \eqref{eq:mbqc_corr} can be expressed as the expectation value of a sum of stabilizers evaluated on the resource state $\rho$. For the state fidelity, each stabilizer appears with an equal coefficient, while this is not the case for MBQC fidelity, implying that certain stabilizers are more crucial to the performance of MBQC than others. To quantify the relationship between $F_S(\rho)$ and $\overbar{F}_\MBQC(\rho)$, we use the theory of quantum state verification \cite{pallister2018optimal,zhu2019general,kliesch2021theory}. In this context, $\Omega$ is a positive semi-definite operator whose largest eigenvalue has a unique eigenstate, namely $\ket{\mathcal{S}}$. Therefore, a large expectation value $\tr(\rho \Omega)$ implies that $\rho$ is close to $\ketbra{\mathcal{S}}$. The precise relationship is \cite{zhu2019general},
\begin{equation} \label{eq:fidbound}
    (1-\tau) F_S(\rho) + \tau \leq \tr \rho \Omega \leq (1-\beta) F_S(\rho) + \beta
\end{equation}
where $\beta$ and $\tau$ are the second largest and smallest eigenvalues of $\Omega$, respectively. In Appendix \ref{app:spectrum}, we show that $\tau = 0$ for all resource states $\ket{\mathcal{S}}$.
Using this and Eq.~\eqref{eq:mbqc_corr}, we can rewrite Eq.~\eqref{eq:fidbound} as,
\begin{equation} \label{eq:final_bound}
    \nu(1-F_S(\rho)) \leq 1-\overbar{F}_\MBQC(\rho) \leq 1-F_S(\rho),
\end{equation}
where $\nu = 1-\beta < 1$ is the spectral gap of $\Omega$. Therefore, the state infidelity directly upper bounds the average MBQC infidelity, greatly improving over the bound based on trace distance.

The bounds in Eq.~\eqref{eq:final_bound} are tight: the ideal resource state has $F_S=\overbar{F}_\MBQC = 1$, saturating the upper bound, while the eigenvector of $\Omega$ with second-largest eigenvalue has $F_S = 0$ and $\overbar{F}_\MBQC = \beta$, saturating the lower bound. The value of $\nu$ depends on the resource state $\ket{\mathcal{S}}$. However, for certain families of resource states of increasing size, we find that $\nu$ is independent of system size. For example, in Appendix \ref{app:1dcluster} we prove that $\nu = 1/4$ for 1D cluster states of arbitrary length $N>2$, and we numerically find that $1/4\leq \nu\leq 1/2$ for 2D cluster states \cite{raussendorf2003measurement} of all sizes $n\times m$ with $n,m\geq 2$ and $nm\leq 20$. This demonstrates the existence of resource states which are orthogonal to the ideal resource states, yet still have a finite value of average MBQC fidelity even in the thermodynamic limit, meaning that infidelity can overestimate average MBQC infidelity by a constant multiplicative factor.




\vspace{3mm}
\textit{Efficient estimation of $\overbar{{F}}_\MBQC$.}---It is well-known that the expression in Eq.~\eqref{eq:state_fidelity} leads to an efficient method to measure state fidelity via direct fidelity estimation, wherein one measures randomly sampled stabilizers on a number of copies of $\rho$ that is independent of system size (for certain well-conditioned classes of states including stabilizer states) \cite{Flammia2011,kliesch2021theory}. Similarly, Eqs.~\eqref{eq:mbqc_corr} and \eqref{eq:omega} can be used to derive an efficient method to measure average MBQC fidelity, where elements of $\mathcal{G}^{XY}$ are now sampled according to the distribution defined by their coefficients in Eq.~\eqref{eq:omega}. Here, we describe an efficient algorithm that accomplishes this sampling.

\begin{figure}[t]
    \centering
    \includegraphics[width=\linewidth]{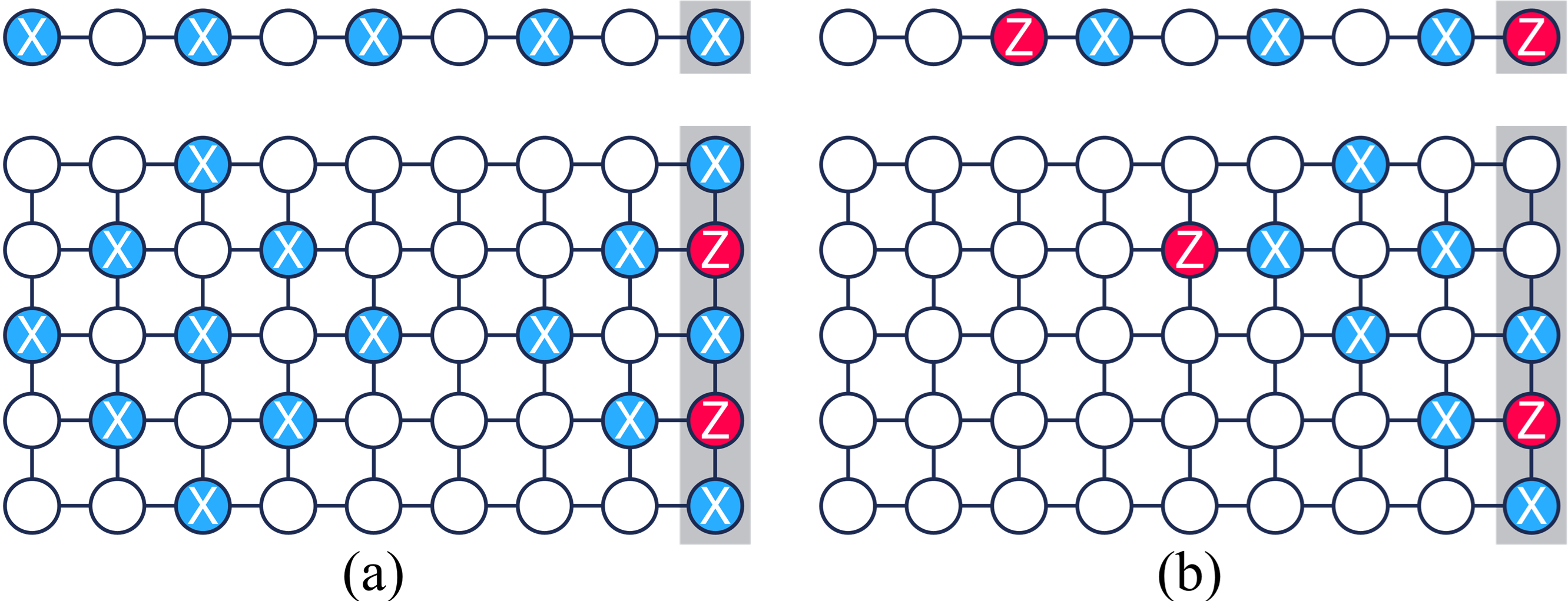}
    \caption{Examples of $T$-stabilizers (left) and $R$-stabilizers (right) for 1D cluster (top) and 2D cluster state with open boundary conditions \cite{raussendorf2003measurement} (bottom). Output qubits are highlighted in the last column.}
    \label{fig:mbqc_cones}
\end{figure}

First, we observe that the flow conditions lead to a useful decomposition of the stabilizer group $\mathcal{G}$. We have already defined a stabilizer $Z_iR_i$ for each qubit $i\in\notouts$, all of which are independent from each other. We call these $R$-stabilizers. Since $\mathcal{G}$ is a complete stabilizer group, we can find $N-|\notouts| = |\outs|$ elements of $\mathcal{G}$ which are mutually independent and independent from all $R$-stabilizers. For each of these elements, we can multiply judiciously by $R$-stabilizers to push all $Z$'s to the output qubits while retaining independence. The resulting stabilizers, which we call $T$-stabilizers and label as $T_i$ for $i\in\mathcal{O}$, therefore act only as $I$ or $X$ on all measured qubits. Together, the $R$- and $T$-stabilizers generate the whole stabilizer group $\mathcal{G}$. Examples of each kind of stabilizer are shown in Fig.~\ref{fig:mbqc_cones} for the 1D and 2D cluster states.

\begin{figure}
    \centering
    \includegraphics[width=\linewidth]{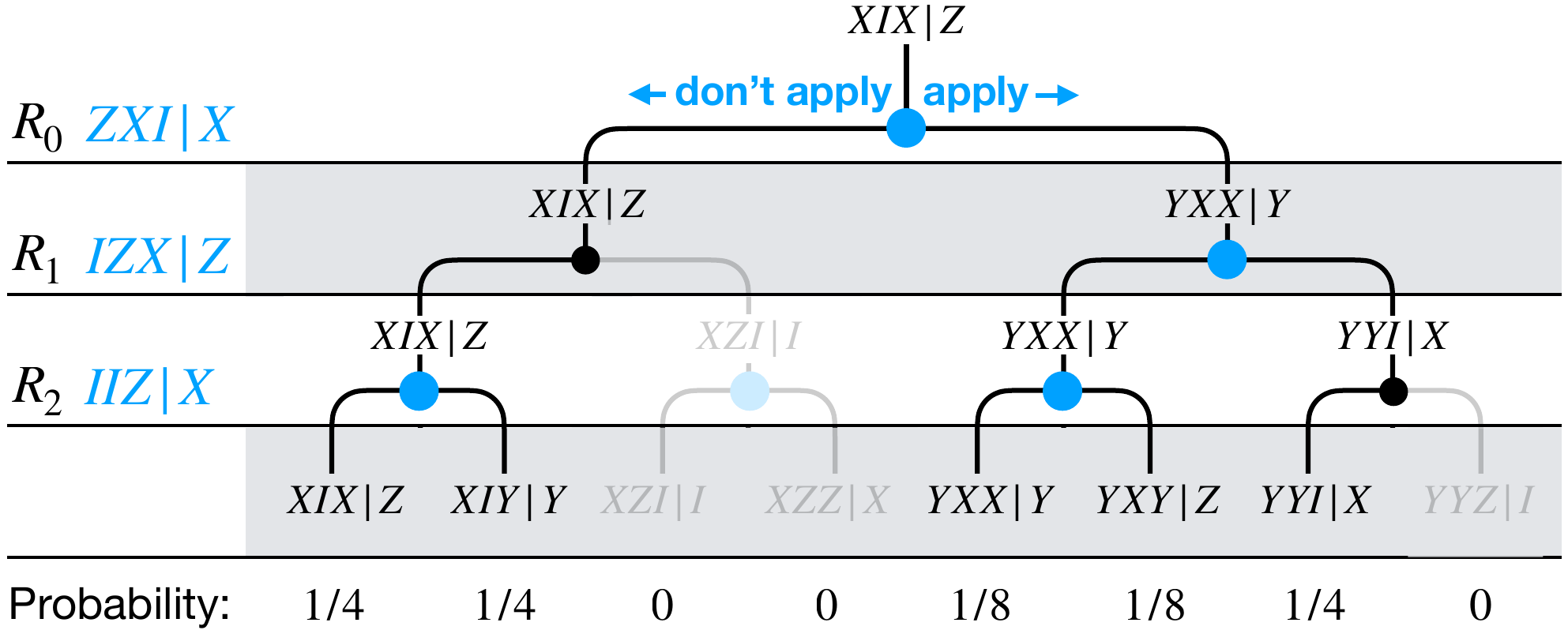}
    \caption{Demonstration of the sampling algorithm for a 1D cluster state of length $N=4$. Each fork in the tree corresponds to multiplying by the corresponding $R$-stabilizer or not. Greyed-out paths in the $i$-th step are not allowed because the stabilizer at that point did not act as $X$ on qubit $i$. Sampling starts at the top of the tree and navigates to the bottom, taking each allowed path with equal probability. This shows one branch of the sampling algorithm corresponding to the $T$-stabilizer $XIXZ$; the only other branch in this case contains only the trivial identity stabilizer. Both branches are then weighted by a probability of $1/2^{|\notouts|} = 1/2$. The vertical bars in each stabilizer represent the division between measured and output qubits.}
    \label{fig:tree}
\end{figure}

The sampling algorithm starts by initializing a stabilizer $g$ as a random element of all $2^{|\outs|}$ products of the $T$-stabilizers. Next, we go through the qubits in temporal order. Whenever we encounter a qubit $i$ on which $g$ acts as $X$, we do nothing with probability $1/2$ and update $g\mapsto gZ_iR_i$ with probability $1/2$. Doing this sequentially for all measured qubits gives the final sampled stabilizer $g$. This algorithm can be viewed as navigating a tree, as shown in Fig.~\ref{fig:tree}. Let us now explain why this algorithm achieves the desired sampling task. First, it is clear that $g\in \mathcal{G}^{XY}$ since the initial product of $T$-stabilizers belongs to $\mathcal{G}^{XY}$ and each subsequent multiplication by an $R$-stabilizer preserves this since the $Z_i$ factor in the $R$-stabilizer changes $X_i$ to $Y_i$ and the $R_i$ factor acts as $I$ or $X$ on all measured qubits. Furthermore, any element of $\mathcal{G}^{XY}$ can be obtained by the above procedure, which can be seen by starting with a given element of $\mathcal{G}^{XY}$ and running the procedure in reverse, multiplying by $R$-stabilizers in reverse temporal order until we are left with a product of $T$-stabilizers. Finally, the probability of a given stabilizer $g$ being sampled is equal to $2^{-|\mathcal{O}|}$ (coming from the initial uniform choice of product of $T$-stabilizers) multiplied by $2^{-w_{\notouts}(g)}$. This latter factor comes from the fact that, for every qubit $i\in\notouts$ on which $g$ acts as $X$ or $Y$, the sampling procedure would have split into two equally probabilistic paths, giving a factor of $1/2$. By definition, there are $w_{\notouts}(g)$ such qubits. Importantly, due to the fact that we move through qubits according to the temporal order $\prec$, multiplying $g$ by an $R$-stabilizer will never alter how $g$ acts on qubits that we already visited in the procedure.

The above algorithm provides a way to efficiently sample elements of $\mathcal{G}^{XY}$ according to the distribution defined by Eq.~\eqref{eq:omega}. Armed with this, the observable $\Omega$, and therefore $\overbar{F}_\MBQC$, can be efficiently estimated using the procedure described in \cite{kliesch2021theory}. To estimate $\Omega$ to additive precision $\epsilon$ with confidence $1-\delta$, one needs to sample and measure $m$ stabilizers where \cite{kliesch2021theory},
\begin{equation}
    m = \frac{2}{\epsilon^2}\ln \frac{2}{\delta}.
\end{equation}
Notably, this sampling overhead is independent of the size of the resource state, and therefore of the size of the implemented circuit in MBQC.

\textit{Conclusion.}---We defined a new metric for benchmarking MBQC resource states, the average MBQC fidelity, and showed how it can be directly estimated using an efficient sampling procedure. This metric is closely related to the fidelity of the noisy resource state with the ideal resource state, providing the most direct guarantee of MBQC performance based on state fidelity to date. However, state fidelity can also greatly underestimate average MBQC fidelity, so measuring the latter directly is preferable (and equally sample-efficient). A natural direction of future research is to evaluate average MBQC fidelity for various resource states subject to various noise models in order to find error-resilient schemes of MBQC.

We are grateful to Y.-H. Chen and K. Mayer for providing comments and reviewing the manuscript. We also thank B. Fefferman, A. Gorshkov, M. Gullans, and D. Hangleiter for helpful comments. 




\bibliography{biblio.bib}

\appendix


\section{Computing $\Omega$ for general resource states} \label{app:omega}

Here we evaluate the integral in Eq.~\eqref{eq:omega_theta} to obtain Eq.~\eqref{eq:omega}. First, we have,
\begin{widetext}
\begin{equation} \label{eq:gamma_s_gamma}
\begin{aligned}
   \Gamma(\bm{\theta})^\dagger \ketbra{\bm{s}} \Gamma(\bm{\theta})  &= \left({\prod_{i\in \notouts}} R_i^{s_i}\right)^\dagger \left({\prod_{i\in \notouts}}   U_i(\theta_i)  \right)^\dagger \ketbra{\bm{s}} \left(\prod_{i\in \notouts}     U_i(\theta_i)\right) \left({\prod_{i\in \notouts}} R_i^{s_i}\right) = \left({\prod_{i\in \notouts}}   U_i(\theta'_i)  \right)^\dagger \ketbra{\bm{s}} \left(\prod_{i\in \notouts}   U_i(\theta'_i)\right).
\end{aligned}
\end{equation}
To get the second equality, we first commute the products of $R_i$, which are Pauli operators, past the $U(\theta)$ gates. This changes $\theta_i$ to $\theta'_i = \pm \theta_i$ where the sign depends on $\bm{s}$. Later, we will integrate each $\theta_i$ such that this sign change will have no effect, and we will drop it from here on. Since each product of $R_i$ acts as $I$ or $X$ on all measured qubits, commuting it past the $H$ in $U(\theta)$ makes it act as $I$ or $Z$, so it is annihilated by the projector $\ketbra{\bm{s}}$.

\end{widetext}

Now, if we write,
\begin{equation}
    \sum_{\bm{s}} \ketbra{\bm{s}}\rho\ketbra{\bm{s}} = \left(\prod_{i\in\notouts} \mathcal{E}^Z_i\right)(\rho),
\end{equation}
where we defined the fully dephasing channel for any Pauli $P=X,Y,Z$,
\begin{equation}
    \mathcal{E}^P(\rho) = \ketbra{0}\rho\ketbra{0} + \ketbra{1}\rho\ketbra{1} = \frac{1}{2}\rho + \frac{1}{2}P \rho P
\end{equation}
then we can plug Eq.~\eqref{eq:gamma_s_gamma} into Eq.~\eqref{eq:omega_theta} to obtain,
\begin{equation} \label{eq:omega_from_exy}
         \Omega = 2^{M} \left(\prod_{i\in\notouts} \mathcal{E}^{XY}_i\right) \left(\ketbra{\mathcal{S}}\right),
\end{equation}
where,
\begin{equation}
    \mathcal{E}^{XY}(\rho) =  \frac{1}{2\pi}\int_{\theta=0}^{2\pi}d\theta\, U(\theta)^\dagger \left(  \mathcal{E}^Z \left(U(\theta)\rho U(\theta)^\dagger  \right) \right)U(\theta).
\end{equation}
Note that we have implicitly performed a change of variables to map $\theta'_i$ back to $\theta_i$ in the integral, for each value of $\bm{s}$.

Next, we compute the integral over $\theta$. Expanding out the dephasing channel, we have,
\begin{equation} \label{eq:exy_int}
\begin{aligned}
    \mathcal{E}^{XY}(\rho) &=  \frac{1}{2}P + \frac{1}{4\pi}\int_{\theta=0}^{2\pi}d\theta\, U(\theta)^\dagger ZU(\theta)\rho U(\theta)^\dagger Z U(\theta) \\
    &= \frac{1}{2}P + X\left(\frac{1}{4\pi}\int_{\theta=0}^{2\pi}d\theta\,  e^{2i\theta Z}\rho e^{-2i\theta Z}\right) X.
\end{aligned}
\end{equation}
To finish, we derive the action of $\mathcal{E}^{XY}$ on Pauli operators, which suffices to define it on all operators since they form a basis. For $\rho = I,X,Y,Z$, a direct computation shows that the remaining integral evaluates to $I,0,0,Z$, respectively. Therefore, we have,
\begin{alignat*}{2} \label{eq:exy}
    \mathcal{E}^{XY}(\id) &= \id \quad
    && \mathcal{E}^{XY}(Z) = 0 \\
    \mathcal{E}^{XY}(X) &= \frac{1}{2}X \quad
    && \mathcal{E}^{XY}(Y) = \frac{1}{2}Y
\end{alignat*}

Finally, to obtain Eq.~\eqref{eq:omega}, we first expand $\ketbra{\mathcal{S}}$ in terms of stabilizers to get,
\begin{equation}
    \Omega = \frac{1}{2^{|\mathcal{O}|}} \sum_{g\in\mathcal{G}} \left(\prod_{i\in\notouts} \mathcal{E}^{XY}_i\right) (g).
\end{equation}
Applying the channels $\mathcal{E}^{XY}_i$ annihilates $g$ if $g$ acts as $Z$ on any measured qubit. It also adds a factor of $1/2$ for each measured qubit on which $g$ acts as $X$ or $Y$. Together, these two actions lead to the stated result in Eq.~\eqref{eq:omega}.

\remark{
The same result can be obtained even if each $\theta_i$ is summed over an appropriate discrete set $\Theta$ of angles, rather than integrated from $0$ to $2\pi$. First, the set $\Theta$ must be invariant under reflections $\Theta \rightarrow -\Theta$ such that the change of variables $\theta'_i\rightarrow \theta_i$ is still possible. Second, in order to ensure that summing over $\Theta$ gives the same result as the integral in Eq.~\eqref{eq:exy_int}, $\Theta$ must satisfy,
\begin{equation}
    \sum_{\theta\in\Theta} e^{4i\theta} = 0.
\end{equation}
The smallest set satisfying both of these conditions is $\Theta = \{0,\pi/4,\pi/2,3\pi/4\}$. This set corresponds to randomly selected each measurement basis from the set $\{\pm X, \pm Y$\}. In MBQC, such Pauli measurements correspond to Clifford gates in the simulated unitary circuit.
}

\remark{
It is easy to generalize the above proof to the scenario where some qubits are deterministically measured in the $X$ or $Y$ basis. Define a map $\mu:\notouts \rightarrow \{X,Y,XY\}$ such that $\mu(i)=X$ ($Y$) if the qubit $i$ is always measured in the $X$ ($Y$) basis, and $\mu(i) =XY$ if the basis is averaged over all angles.
Then a slight modification of the above proof yields,
\begin{equation}
    \Omega = \frac{1}{2^{|\mathcal{O}|}} \sum_{g\in\mathcal{G}} \left(\prod_{i\in\notouts} \mathcal{E}^{\mu(i)}_i\right) (g).
\end{equation}
Similar to the above analysis, this product of channels annihilates all stabilizers $g$ which act as $Z$ or $Y$ on any qubit $i$ with $\mu(i) = X$, as $Z$ or $X$ on qubits with $\mu(i) = Y$, or as $Z$ on qubits with $\mu(i) = XY$. Finally, it adds a factor of $\frac{1}{2}$ for each qubit with $\mu(i) = XY$ on which $g$ acts as $X$ or $Y$.
}

\remark{
Finally, we can write a useful alternative representation of $\Omega$ in terms of the procedure to sample elements of $\mathcal{G}^{XY}$ described in the main text. If we express that procedure as an equation, we have, 
\begin{equation} \label{eq:omega_algo}
    \Omega = \frac{1}{2^{|\outs|}}\sum_{t \in\mathcal{T}} \left( \prod_{i\in\notouts} \Lambda_i\right)(t) := \frac{1}{2^{|\outs|}}\sum_{t \in\mathcal{T}} \Lambda_{\notouts}(t)
\end{equation}
where, $\mathcal{T}\subset \mathcal{G}$ is the subgroup generated by all $T$-stabilizers (see main text), the product is taken in temporal order as usual, and the the superoperator $\Lambda_i$ is defined by its action on any Pauli string $P$,
\begin{equation}
    \Lambda_i(P) = \begin{cases}
        \hfil P & \mathrm{if} \quad [P,Z_i] = 0 \\
        \frac{1}{2}(P + PZ_iR_i) & \mathrm{if} \quad \{P,Z_i\} = 0
    \end{cases}
\end{equation}
The form in Eq.~\eqref{eq:omega_algo} represents $\Omega$ as a sum of all operators that come out of the sampling procedure, weighted by their probability. Observe that, if $g$ is a stabilizer, then $\Lambda_{\notouts}(g) $ is a sum of stabilizers with coefficients that sum to 1. Since there are $2^{|\ins|} = 2^{|\outs|}$ such terms terms in Eq.~\eqref{eq:omega_algo}, each weighted by $2^{|\outs|}$, the total sum of all coefficients in $\Omega$ is equal to 1.
}

\section{Spectrum of $\Omega$} \label{app:spectrum}

Here we prove various properties of the spectrum of $\Omega$. First, we observe that the spectrum of $\Omega$ is contained in $[0,1]$, which is most easily seen by the fact that $\tr \rho\Omega$ is an average of state fidelities, each of which is contained in $[0,1]$. Second, since $\overbar{F}_\MBQC(\ketbra{\mathcal{S}}) = 1$, we must have $\Omega\ket{\mathcal{S}} = \ket{\mathcal{S}}$. In the rest of this section, we aim to prove (1) $\ket{\mathcal{S}}$ is the unique eigenstate of $\Omega$ with eigenvalue 1 and (2) the minimum eigenvalue of $\Omega$ is equal to 0. 

\subsection{Largest eigenvalue}

We first show that $\ket{S}$ is the unique eigenstate with eigenvalue 1. This requires understanding some structure of the set $\mathcal{G}^{XY}$ of stabilizers that act as $I$, $X$, or $Y$ on all measured qubits.
Since $\Omega\ket{\mathcal{S}} = \ket{\mathcal{S}}$, and $g\ket{\mathcal{S}} = \ket{\mathcal{S}}$ for all $g\in\mathcal{G}$, it must be the case that all coefficients in Eq.~\eqref{eq:omega} sum to 1. Then, if another state $\ket{\psi}$ also satisfies $\Omega\ket{\psi} = \ket{\psi}$, we have $g\ket{\psi} = \ket{\psi}$ for all $g\in\mathcal{G}^{XY}$. Therefore, if $\mathcal{G}^{XY}$ generates all of $\mathcal{G}$, then $\Omega\ket{\psi} = \ket{\psi}$ implies that $g\ket{\psi} = \ket{\psi}$ for all $g\in\mathcal{G}$, which implies that $\ket{\psi} = \ket{\mathcal{S}}$, as desired.

To prove that this is the case, it is sufficient to show how to obtain any $T$-stabilizer or $R$-stabilizer as a product of elements of $\mathcal{G}^{XY}$ as they generate all of $\mathcal{G}$. This case of $T$-stabilizers is trivial, since they already act as $I$ or $X$ on all measured qubits by definition and therefore belong to $\mathcal{G}^{XY}$. To get an arbitrary $R$-stabilizer $Z_iR_i$, we will identify an element of $g\in \mathcal{G}^{XY}$ that acts as $X$ or $Y$ on qubit $i$. Then, $gZ_iR_i$ will also belong to $\mathcal{G}^{XY}$, such that $Z_iR_i = (g)(gZ_iR_i)$ is a product of elements of $\mathcal{G}^{XY}$. 

Let us show how to find such an element $g$ for all $i$. First, for every qubit $i$, such an element $g$ must exist amongst the $T$-stabilizers and $R$-stabilizers since otherwise $Z_i$ would commute with all of $\mathcal{G}$, which implies that qubit $i$ is completely disentangled, and this is not allowed by our assumptions. The question is then whether such a $g$ can also be found in $\mathcal{G}^{XY}$.

We use induction on the qubits according to the temporal order. First, note that the only stabilizer generators which can possibly act as $X$ or $Y$ on the first qubit $i=1$ in the temporal order are the $T$-stabilizers, which are contained in $\mathcal{G}^{XY}$. Therefore, there exists an element of $\mathcal{G}^{XY}$ that acts as $X$ on qubit $i=1$. Now, suppose we have shown that, for all $i\prec j$, there exists an element $g_i\in \mathcal{G}^{XY}$ that acts as $X$ on qubit $i$. We need to show the same for qubit $j$. Let $g$ be a generator that acts as $X$ on qubit $j$, which we argued above must exist. If $g$ is an $T$-stabilizer then we are done since $T_i\in\mathcal{G}^{XY}$. Otherwise, $g$ is a $R$-stabilizer $Z_iR_i$ stabilizer for some $i\prec j$. Then, by the inductive assumption, we can multiply by the element $g_i$ to obtain the element $g_j = g_iZ_iR_i$ which acts as $X$ on qubit $j$ and belongs to $\mathcal{G}^{XY}$. This concludes the proof.

\subsection{Smallest eigenvalue}

To find the minimum eigenvalue of $\Omega$, we use the alternative representation in Eq.~\eqref{eq:omega_algo}.
Since $\Lambda_{\notouts}(g) $ is a sum of stabilizers with coefficients that sum to 1, we have,
\begin{equation}
    \Lambda_{\notouts}(t) \ket{S} = 1
\end{equation}
for all $t\in \mathcal{T}$.
Now consider the ``excited state'' $\ket{\mathcal{S}_{k}}$ which is defined by the eigenvalue relations $T_i\ket{\mathcal{S}_{k}} = (-1)^{\delta_{i,k}}\ket{\mathcal{S}_{k}}$ and $Z_iR_i\ket{\mathcal{S}_{k}} = \ket{\mathcal{S}_{k}}$ for all $i$. Then, for every $t\in\mathcal{T}$, we have $t\ket{\mathcal{S}_{k}} = \pm \ket{\mathcal{S}_{k}}$, where the sign is negative for exactly half of the elements of $\mathcal{T}$ (since half of the elements contain $T_i$ as a factor). Similarly, we have $\Lambda_{\notouts}(t) = \pm \ket{\mathcal{S}_{k}}$ with the same sign as above, since $Z_iR_i\ket{\mathcal{S}_{k}} = \ket{\mathcal{S}_{k}}$ for all $i$. Therefore, we find that 
\begin{equation}
    \Omega \ket{\mathcal{S}_{k}} =  \frac{1}{2^{|\outs|}}\sum_{t \in\mathcal{T}} \Lambda_{\notouts}(t)\ket{\mathcal{S}_{k}} = \frac{1}{2^{|\outs|}}\sum_{t \in\mathcal{T}} (\pm 1)\ket{\mathcal{S}_{k}} = 0
\end{equation}
where the sum evaluates to 0 due to the equal balance between $\pm 1$ signs. Thus, the states $\ket{\mathcal{S}_{k}}$ all have eigenvalue 0.

Intuitively, if we look at the effective unitary circuit implemented by MBQC, then the states $\ket{\mathcal{S}_{k}}$ result in the same unitary evolution as the ideal resource state $\ket{\mathcal{S}}$, but with an orthogonal initial state. Therefore, the fidelity with the ideal output will always be exactly 0.

\section{1D cluster state example} \label{app:1dcluster}

In this section we study the case of the 1D cluster state in detail. We will derive a recurrence relation for $\Omega$ as a function of number of qubits $N$, writing $\Omega_N$ to be explicit. Then we will use this relation to determine the spectral gap of $\Omega_N$ for all $N$. 

\subsection{Recurrence relation for $\Omega$}

We begin by transforming $\Omega_N$ in a way that will simplify the upcoming analysis. First, we multiply by a factor of 2 and subtract off the component proportional to the identity. Second, we conjugate by the unitary $\overbar{\CZ} = \prod_{i=1}^{N-1} \CZ_{i,i+1}$ where $\CZ = \ketbra{0}\otimes \id + \ketbra{1}\otimes Z$ is the controlled-$Z$ operator. This unitary maps the 1D cluster state to the product state $\ket{++\dots +}$. Together, we get the modified operator,
\begin{equation} \label{eq:omega_tilde_to_omega}
    \tilde{\Omega}_N = \overbar{\CZ} \left(2\Omega_N -  \id \right) \overbar{\CZ}.
\end{equation}
Since $\Omega_N$ is a sum of stabilizers of the cluster state, $\tilde{\Omega}_N$ will be a sum of non-trivial stabilizers of the product state, \textit{i.e.}, non-trivial Pauli strings consisting of $\id$ and $X$ only ($X$-strings). The coefficient of a given $X$-string is the weight of that string on all non-ouput qubits, and all coefficients sum to 1. For example, from Eq.~\eqref{eq:omega23}, we can compute,
\begin{equation} \label{eq:omegatilde23}
\begin{aligned}
    \tilde{\Omega}_2 &= \frac{1}{2}XI + \frac{1}{2}XX \\
    \tilde{\Omega}_3 &= \frac{1}{2}XIX + \frac{1}{4}XXX + \frac{1}{4}XXI.
\end{aligned}
\end{equation}
The $X$-strings appearing in $\tilde{\Omega}_N$ are those which are mapped to elements of $\mathcal{G}^{XY}$ upon conjugation by $\overbar{\CZ}$. Noting that $\overbar{\CZ} X_i \overbar{\CZ} = Z_{i-1}X_i Z_{i+1}$ (with appropriate modifications for $i=1,N$), and calling an $X$-string ``empty'' on site $i$ if it acts as $I$ on that site and ``filled'' if it acts as $X$, it is not hard to see that the allowed $X$-strings are those with no neighboring empty sites, and with the site $i=1$ being filled.

Now, suppose we take an $X$-string appearing in $\tilde{\Omega}_N$ and we remove the first (leftmost) qubit. If the resulting truncated string is filled on the new leftmost qubit, then it appears in $\tilde{\Omega}_{N-1}$. If it is empty on the new leftmost qubit, then it must be filled on the next qubit since there are no neighboring empty sites, such that truncating the leftmost qubit again results in a stabilizer that appears in $\tilde{\Omega}_{N-2}$. The converse works as well: appending an $X$ ($XI$) to the left of any stabilizer contained in $\tilde{\Omega}_{N-1}$ ($\tilde{\Omega}_{N-2}$) yields a stabilizer contained in $\Omega_N$.

These observations lead us to the recurrence relation,
\begin{equation} \label{eq:recurrence}
    \tilde{\Omega}_N = \frac{1}{2} X\otimes \tilde{\Omega}_{N-1} + \frac{1}{2} XI\otimes \tilde{\Omega}_{N-2},
\end{equation}
which, along with the Eq.~\eqref{eq:omegatilde23}, allows us to recursively define all $\tilde{\Omega}_N$ for $N>3$. Even without explicitly determining the operators, this relation will allow us to extract their most relevant properties.


\subsection{Spectrum of $\Omega$ for 1D cluster state}

Now we use the recurrence relation \eqref{eq:recurrence} to understand the spectrum of $\Omega_N$ for arbitrary $N$.
Specifically, we will prove that $\beta = 3/4$ (second largest eigenvalue) for all $N>2$. From Eq.\eqref{eq:omega_tilde_to_omega}, we see that the eigenvalues $\tilde{\lambda}_i$ of $\tilde{\Omega}_N$ are related to the eigenvalues $\lambda_i$ of $\Omega_N$ by $\tilde{\lambda}_i = 2\lambda_i - 1$. Therefore, we need to show that the second largest eigevanlue $\tilde{\beta}$ of $\tilde{\Omega}_N$ equals 1/2 for all $N>2$. 

Note that the eigenvectors of $\tilde{\Omega}$ can be taken to be product states of the form $\ket{\pm \pm\dots \pm}$. Write $\ket{++\dots +} = \ket{+_N}$. We have shown that the cluster state is an eigenvector of $\Omega_N$ with eigenvalue 1, so $\ket{+_N}$ must be an eigenstate of $\tilde{\Omega}_N$ with eigenvalue 1. We also showed that every stabilizer contained in $\tilde{\Omega}_N$ acts as $X$ on the leftmost qubit, so the spectrum is symmetric about 0, in particular we have $Z_1\tilde{\Omega}_N = -\tilde{\Omega}_N Z_1$, such that $Z_1\ket{+_N}$ is an eigenstate with eigenvalue $-1$.



Now, we prove that the state $Z_1Z_2Z_3\ket{+_N}$ has eigenvalue $1/2$ for all $N\geq 3$. Using Eq.~\eqref{eq:recurrence} twice, we can derive,
\begin{equation}
    \tilde{\Omega}_N = \left(\frac{1}{2} XI +\frac{1}{4}XX\right)\otimes  \tilde{\Omega}_{N-2} + \frac{1}{4}XXI\otimes \tilde{\Omega}_{N-3}
\end{equation}
Now, using the facts that $\tilde{\Omega}_k\ket{+_k} = \ket{+_k}$ and $\tilde{\Omega}_k Z_1 \ket{+_k} = -Z_1 \ket{+_k}$ for all $k\geq 2$, we have,
\begin{equation}
\begin{aligned}
    \tilde{\Omega}_N Z_1Z_2Z_3\ket{+_N} &= \left(\frac{1}{2} - \frac{1}{4}  + \frac{1}{4}\right) Z_1Z_2Z_3\ket{+_N} \\
    &=\frac{1}{2} Z_1Z_2Z_3\ket{+_N}
\end{aligned}
\end{equation}

Now, we prove that every eigenstate $\ket{\psi}$ other than $\ket{+_N}$ has eigenvalue $\lambda \leq 1/2$ for $N\geq 3$. We split the analysis into several cases. 

(i) $\ket{\psi} = Z_1 \ket{+_N}$. We have already argued that this case gives eigenvalue $\lambda=-1$.

(ii) $\ket{\psi} = Z_2 \ket{+_N}$. Using the facts that $\tilde{\Omega}_k\ket{+_k} = \ket{+_k}$ and that $Z_2$ anticommutes (commutes) with the first (second) term in Eq.~\eqref{eq:recurrence}, we get $\lambda = -\frac{1}{2}+\frac{1}{2} = 0$.

(iii) $\ket{\psi} = Z_1Z_2 \ket{+_N}$. Similar to the above case, except now $Z_1Z_2$ anticommutes (commutes) with first (second) term in Eq.~\eqref{eq:recurrence}, giving $\lambda = \frac{1}{2} - \frac{1}{2} = 0$.

(iv) $\ket{\psi} = \ket{ab}\otimes \ket{\phi}$ where $a,b\in\{+,-\}$ and $\ket{\phi}$ is some product state that is not equal to $\ket{+_{N-2}}$. This case is covered by induction. Assume that, for all $3\leq k < N$, the maximum eigenvalue of $\tilde{\Omega}_k$ for all eigenstates other than $\ket{+_k}$ is $\leq 1/2$. Because $\tilde{\Omega}_k$ anticommutes with $Z_1$, the spectrum must be symmetric about 0, so this assumption also implies that the minimum eigenvalue for all states other than $Z_1\ket{+_k}$ is $\geq -1/2$. The base cases $N=3,4$ can be easily proven by directly computing $\tilde{\Omega}$ and its spectrum. Now consider $N\geq 5$. From Eq.~\eqref{eq:recurrence}, we get,
\begin{equation}
\begin{aligned}
    \tilde{\Omega}_N\ket{\psi} &= \frac{(-1)^a}{2}\ket{a}\otimes \tilde{\Omega}_{N-1} ( \ket{b}\otimes\ket{\phi}) \\
    &+ \frac{(-1)^a}{2} \ket{ab}\otimes \tilde{\Omega}_{N-2} \ket{\phi} \\
    &\leq \left(\frac{1}{2} \times \frac{1}{2}  + \frac{1}{2}\times\frac{1}{2}\right)\ket{\psi}\\
    &=\frac{1}{2}\ket{\psi}
\end{aligned}
\end{equation}
where the inequality follows from the inductive assumptions which imply,
\begin{equation}
\begin{gathered}
     -\frac{1}{2} \leq (\bra{b} \otimes \bra{\phi}) \tilde{\Omega}_{N-1} ( \ket{b}\otimes\ket{\phi}) \leq \frac{1}{2} \\
     -\frac{1}{2} \leq \bra{\phi} \tilde{\Omega}_{N-2} \ket{\phi} \leq \frac{1}{2}
\end{gathered}
\end{equation}
for all $\ket{\phi} \neq \ket{+_{N-2}}$.
Since cases (i)-(iv) cover all eigenstates not equal to $\ket{+_N}$, this concludes the proof.

\end{document}